\documentclass[12pt,a4paper]{article} 
\usepackage{bbm}
\usepackage[dvips]{graphicx}

\long\def\symbolfootnote[#1]#2{\begingroup\def\thefootnote{\fnsymbol{footnote}}\footnote[#1]{#2}\endgroup} 

\begin{document}
\begin{flushright}
 DESY 09-159\\
 SFB/CPP-09-90\\
\end{flushright}
\begin{center}
\Large
Continuum-limit scaling of overlap fermions\\
as valence quarks

\normalsize

\vspace{0.6cm}

KRZYSZTOF CICHY\symbolfootnote[1]{Presented at the 49. Cracow School of Theoretical Physics, Zakopane, Poland, 31 May -- 10 June 2009.}\\
\emph{Adam Mickiewicz University, Faculty of Physics,\\
Umultowska 85, 61-614 Poznan, Poland}\\

\vspace{0.3cm}

GREGORIO HERDOIZA, KARL JANSEN\\
\emph{NIC, DESY, Platanenallee 6, D-15738 Zeuthen, Germany}

\begin{abstract}
\noindent We present the results of a mixed action approach, employing dynamical twisted mass fermions in the sea sector and overlap valence fermions, with the aim of testing the continuum limit scaling behaviour of physical quantities, taking the pion decay constant as an example. To render the computations practical, we impose for this purpose a fixed finite volume with lattice size $L\approx1.3$ fm. We also briefly review the techniques we have used to deal with overlap fermions.
\end{abstract}


\end{center}

  
\section{Introduction}
Lattice QCD is considered to be the most effective way of studying the non-perturbative aspects of the field theory of strong interactions, Quantum Chromodynamics (QCD).
It is a regularization of QCD, which consists in discretizing the relevant degrees of freedom by putting them on a four dimensional lattice with lattice spacing $a$, its inverse being the ultraviolet cutoff of the theory.
With the present generation of supercomputers, large scale dynamical simulations are performed by a number of collaborations using different types of gauge and fermionic lattice actions. However, for some classes of actions, such as overlap fermions, which exactly preserve chiral symmetry, such simulations are extremely demanding\footnote{For instance, the JLQCD and TWQCD collaborations are simulating dynamical overlap fermions, but in their case the global topological sector is kept fixed \cite{jlqcd}.}.
A promising alternative for fully dynamical simulations with overlap fermions is the mixed
action approach, which consists in using a computationally faster
formulation for the fermions in the sea, such as Wilson twisted mass
fermions, while using overlap fermions in the valence sector. In this
way, one avoids the most computer-intensive part of a dynamical
simulation with overlap fermions, but at the same time one profits from having such fermions in the valence sector. In the case of overlap fermions, its exact chiral symmetry gives the benefit of simplifying the operator mixing problem, something which is essential in several types of lattice computations, like the determination of the kaon bag parameter $B_K$.

In this paper, we present the results of a mixed action approach with overlap fermions in the valence sector and Wilson twisted mass fermions in the sea sector\footnote{For an account of earlier stages of the project, see \cite{OVonTM}, \cite{OVonTM2}.}. In Section 2 we briefly review the overlap formulation and the techniques necessary to effectively apply it. Section 3 gives the description of our setup and in Section 4 we present the continuum-limit scaling test of overlap fermions and we discuss the results. Section 5 concludes.

\section{A brief review of overlap fermions}
Since this contribution is written for a more general audience, we provide here a very short review of overlap fermions.
\subsection{The need for overlap fermions}
A very general problem of lattice field theory with fermions is the doubling problem. With a naive fermion discretization, instead of one fermion in the continuum limit, one has 16 fermions in 4 dimensions of spacetime.
As was originally proposed by Wilson \cite{wilson}, the doubling problem can be solved by using the following Wilson-Dirac operator:
\begin{equation}
D_W(m)=\frac{1}{2}\left(\gamma_\mu(\nabla_\mu^*+\nabla_\mu)-ar\nabla_\mu^*\nabla_\mu\right)+m,
\end{equation}
where $\nabla_\mu^*$ ($\nabla_\mu$) is the backward (forward) lattice derivative, $r$ -- Wilson parameter, $m$ -- bare quark mass.
However, the Wilson term in the above Dirac operator explicitly breaks chiral symmetry, which is in accordance with a general theorem proved by Nielsen and Ninomiya \cite{nielsen} back in 1981 that it is impossible to have at the same time for a Dirac operator $D$: locality, translational invariance, no doublers and chiral symmetry, i.e. $\{D(p),\gamma_5\}=0$.
Since Wilson's proposal, much of the effort went into finding a lattice theory
without doublers which preserves the largest possible number of symmetries, and at the same time reaches the continuum limit as fast as possible (which practically means the absence of $\mathcal O(a)$ leading cutoff dependence).

In 1982 (i.e. only one year after establishing the Nielsen-Ninomiya theorem), it was shown by Ginsparg and Wilson \cite{ginsparg} that a remnant of chiral symmetry is present on the lattice without the doubler modes, if the corresponding Dirac operator $D$ obeys an equation now called the Ginsparg-Wilson relation:
\begin{equation}
\gamma_5 D + D\gamma_5 = aD\gamma_5 D.
\end{equation} 
However, the solutions to this equation were not known for many years, until a particularly simple form of a Dirac operator obeying the Ginsparg-Wilson relation was given by Neuberger \cite{neuberger} in 1997.
Neuberger's discretization is now usually referred to as overlap fermions\footnote{For a review of overlap fermions see e.g. \cite{niedermayer}.} and the (massless) overlap Dirac operator is of the form:
\begin{equation}
\label{massless_overlap}
D_{ov}(0)=\frac{1}{a}\Big(1-A(A^\dagger A)^{-1/2}\Big),
\end{equation}
where:
\begin{equation}
A=1-aD_{W}(0).
\end{equation}
It was also found that the overlap Dirac operator is local under very general conditions \cite{hernandez}. 
Moreover, in 1998 L\"uscher \cite{luscher} found that the Ginsparg-Wilson relation leads to a non-standard realization of lattice chiral symmetry. The action is invariant under:
\begin{equation}
\psi\rightarrow e^{i\theta\gamma_5\left(1-\frac{aD}{2}\right)}\psi,
\end{equation}
\begin{equation}
\bar\psi\rightarrow\bar\psi e^{i\theta\gamma_5\left(1-\frac{aD}{2}\right)}.
\end{equation}
Although it is not the standard form of chiral symmetry, it still correctly reproduces the anomaly and protects the fermions from additive mass renormalization and $\mathcal{O}(a)$ lattice artifacts.
The non-standard realization of chiral symmetry means also that the conditions of the Nielsen-Ninomiya theorem do not apply and one can have chiral symmetry (which becomes standard chiral symmetry in the continuum limit) without the doublers.

In order to simulate massive overlap quarks, one uses the following form of the Dirac operator:
\begin{equation}
D_{ov}(m) = \left(1-\frac{am}{2}\right)D_{ov}(0)+m,
\end{equation} 
where $m$ is the bare overlap quark mass and $D_{ov}(0)$ the massless overlap Dirac operator of eq. (\ref{massless_overlap}).

\subsection{Techniques to effectively deal with overlap fermions}
The main disadvantage of overlap fermions is that they are much more costly to simulate
-- by a factor of 30-120 in comparison with maximally twisted mass fermions \cite{chiarappa}. Moreover, this factor increases when approaching the physical pion mass. Therefore, it was essential to develop techniques to effectively deal with overlap fermions. The aim of this subsection is to briefly review them.

\noindent\textbf{Computation of the overlap operator.} The first important thing is to effectively calculate the overlap Dirac operator itself. This is non-trivial, since it involves the inverse of the square root of a matrix. There are several ways to do this, including polynomial approximations, Lanczos based methods and partial fraction expansion. For overviews of these methods, see e.g. \cite{frommer02}, \cite{frommer}. The method that we have chosen is a Chebyshev polynomial approximation, which consists in constructing a polynomial $P_{n,\epsilon}(x)$ of degree $n$, which has an exponential convergence rate in the interval $x\in [\epsilon,1]$. The advantages of using this approximation are the well-controlled exponential fit accuracy and the possibility of having numerically very stable recursion relations which allows for large degrees of the polynomial. To ensure that the Ginsparg-Wilson relation (for massless Dirac operator) is fulfilled to a very high precision, the degree of Chebyshev polynomial $n$ has to satisfy the following condition:
\begin{equation}
 ||X-P_{n,\epsilon}(A^\dagger A) A^\dagger A P_{n,\epsilon}(A^\dagger A) X||^2/||X||^2 < \xi,
\end{equation} 
where $\xi$ has to be a very small number, typically set to $10^{-16}$ to achieve a compromise between good quality of approximation and its cost.
The degree of the polynomial depends on the condition number of the matrix $A^\dagger A$, i.e. the ratio of the highest to the lowest eigenvalue. The lowest eigenvalue can be a very small number and hence the condition number can be prohibitively large, if one constructs the approximation on the interval $[\epsilon,1]$, with $\epsilon$ being the lowest eigenvalue. Fortunately, one can do much better with the following method.

\begin{figure}[t]
\includegraphics[width=0.36\textwidth,angle=270]{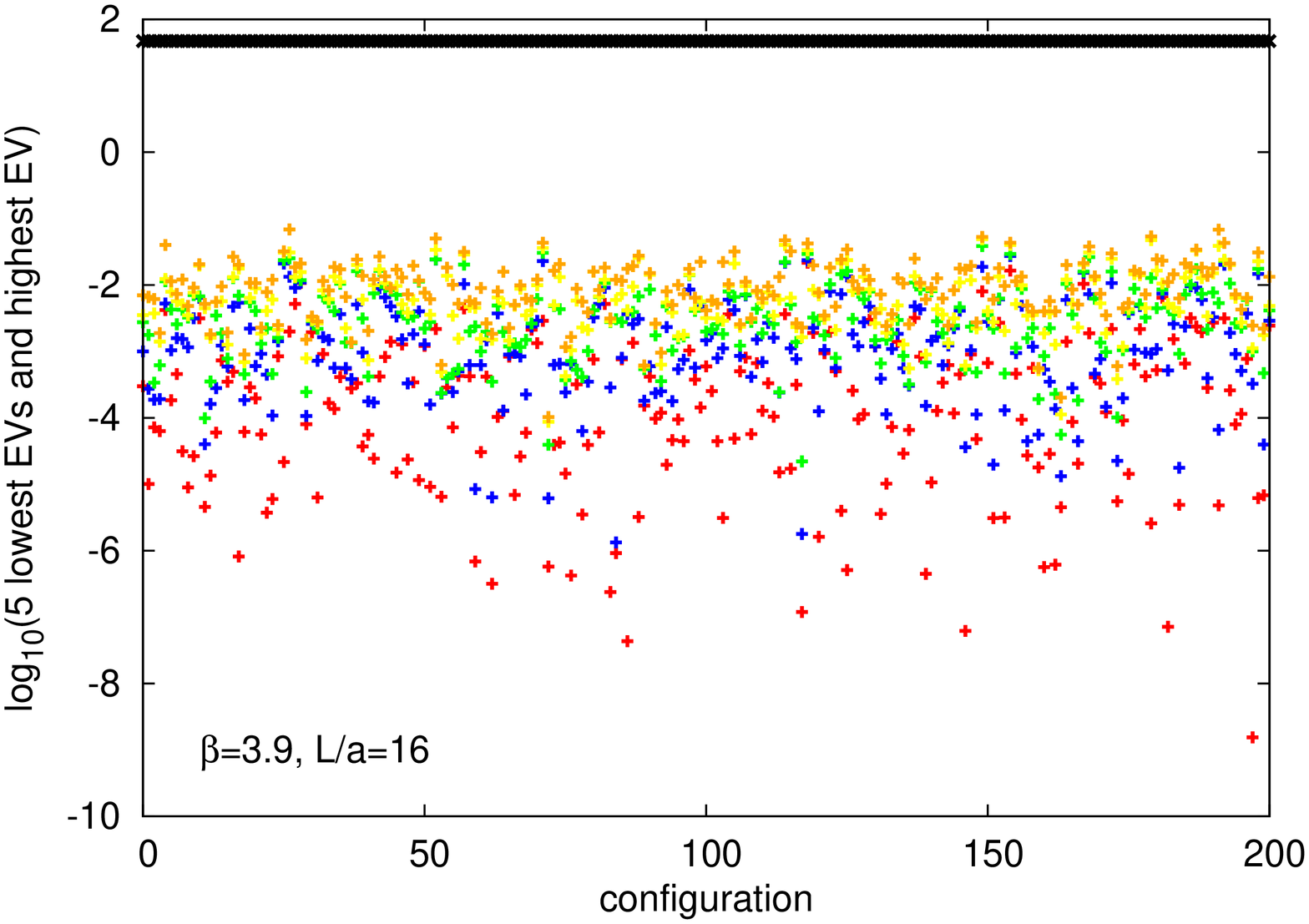}
\includegraphics[width=0.36\textwidth,angle=270]{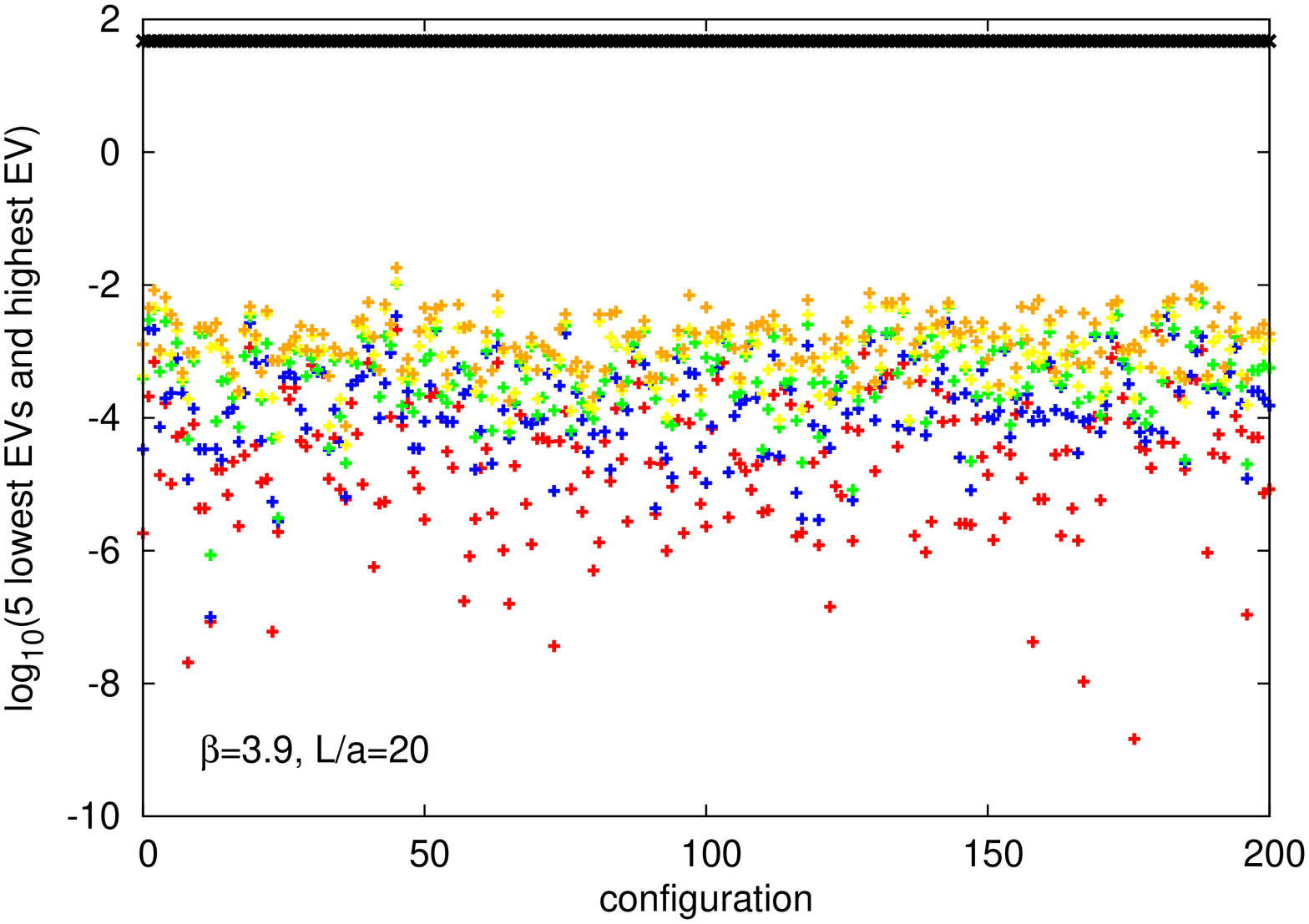}
\includegraphics[width=0.36\textwidth,angle=270]{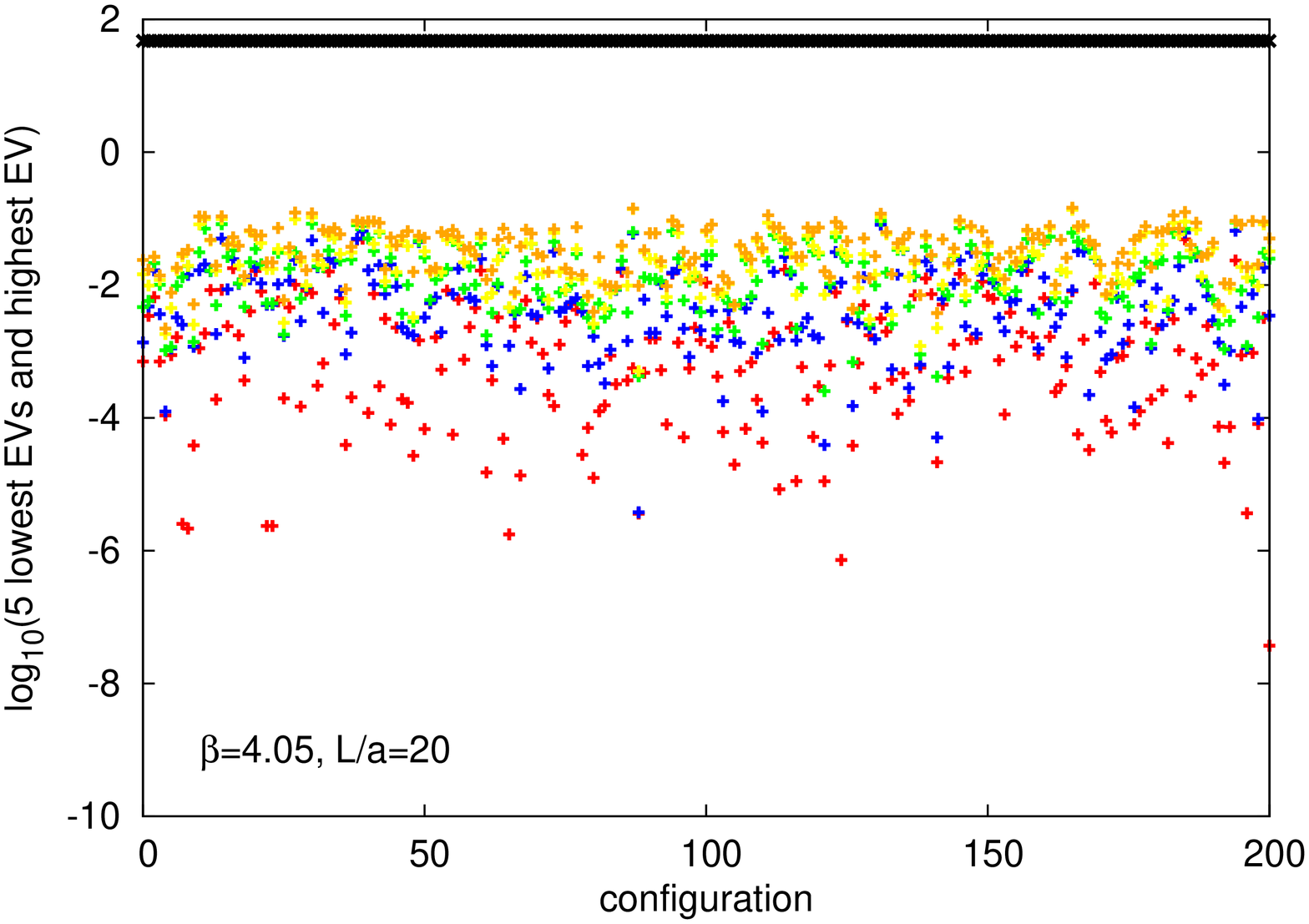}
\includegraphics[width=0.36\textwidth,angle=270]{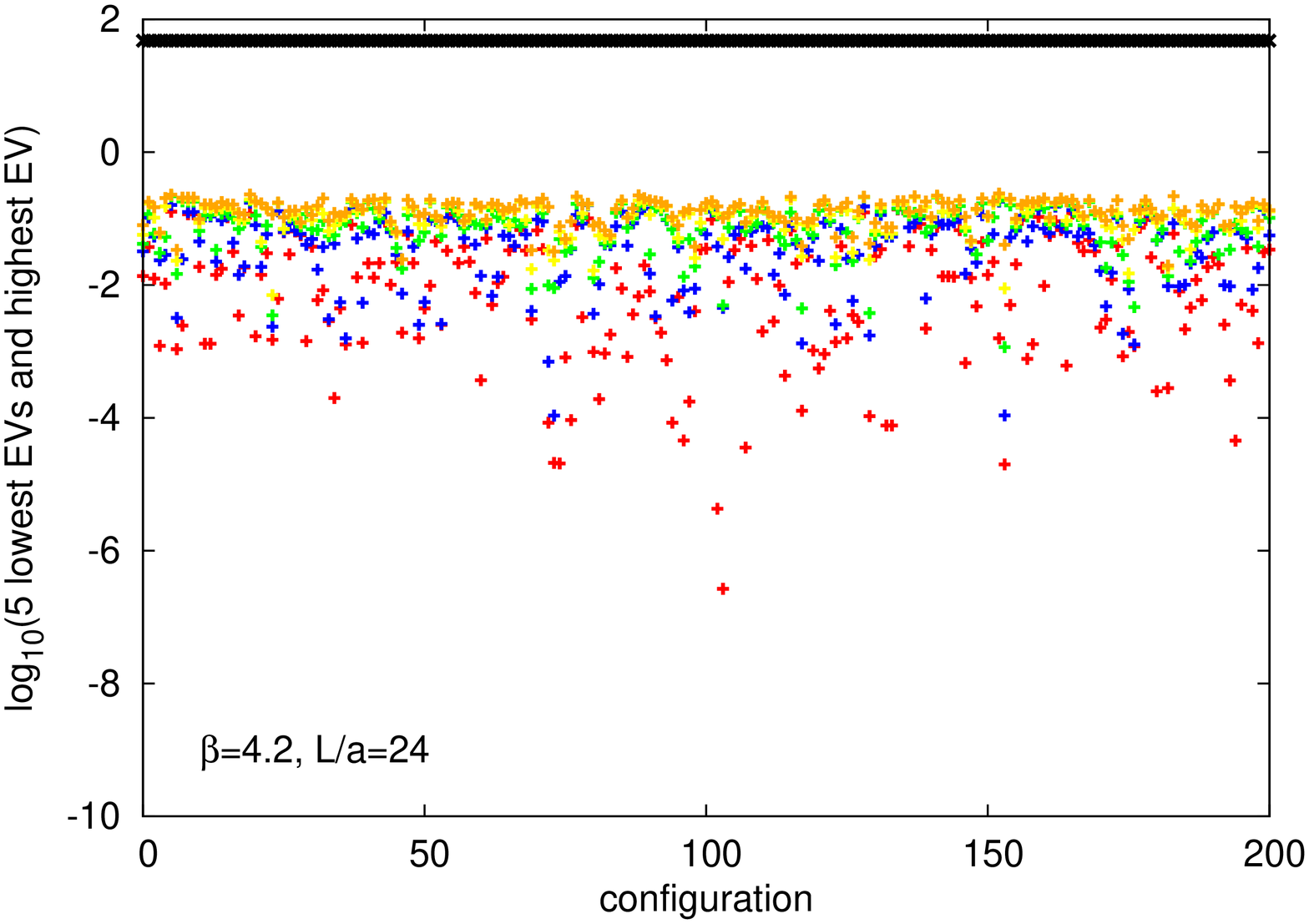}
\caption{5 lowest eigenvalues and the highest eigenvalue for various gauge field ensembles. The lattice spacing is $a\approx0.084$ fm for upper plots, $a\approx0.066$ fm for bottom left and $a\approx0.053$ fm for bottom right plot. \label{EVs}}
\end{figure}

\noindent\textbf{Eigenvalue deflation.} To achieve a considerably smaller degree of Chebyshev polynomial, one should calculate a certain number\footnote{The exact number has to be found empirically and tends to increase with volume.} of the lowest eigenvalues and eigenvectors of $A^\dagger A$ and project them out of this matrix. In this way, the Chebyshev approximation is constructed on the interval $[\epsilon,1]$ with $\epsilon$ equal to the highest of the computed eigenvalues. To illustrate how low the lowest eigenvalues of the matrix $A^\dagger A$ can go, we plotted in Fig. \ref{EVs} the five lowest eigenvalues and the highest eigenvalue for four gauge field ensembles, each with 200 configurations. The general dependence that can be deduced from these plots is that increasing lattice spacing (decreasing $\beta$) moves the spectrum down (eigenvalues in lattice units tend to become smaller) and increases the probability of having very low eigenvalues.

\noindent\textbf{HYP smearing of gauge fields.} Another way to lower the condition number of the matrix $A^\dagger A$ and thus the degree of the Chebyshev polynomial is to perform one iteration of HYP smearing on the gauge fields. This technique was introduced by A. Hasenfratz and F. Knechtli \cite{hyp} and allows to achieve much better convergence of the solver due to improved chiral properties\footnote{Hasenfratz and Knechtli \cite{hyp} remark that fat links lead to an order of magnitude improvement in convergence.}. In comparison with other link fattening methods (e.g. APE smearing), HYP smearing is believed to preserve better the short-distance quantities, because it mixes links from hypercubes attached only to the original link.

\noindent\textbf{SUMR solver with adaptive precision and multiple mass capability.} Having constructed the Dirac operator (with Chebyshev approximation), to find the propagator one has to solve the equation:
\begin{equation}
\label{dirac_invert}
 D_{ov}(m)\psi=\eta,
\end{equation} 
where $\psi$ is the propagator and $\eta$ is the source -- a vector whose choice will be commented on below. To effectively solve this equation, one has to choose the most appropriate solver. Chiarappa et al. \cite{chiarappa} found that in the case of (quenched) overlap and small volume ($12^4$ and $16^4$), the most effective solver is the chiral conjugate gradient algorithm, with the SUMR solver just behind\footnote{SUMR = Shifted Unitary Minimal Residual.}. However, the former algorithm can only be used for exact overlap operator, which means that the polynomial approximation would lead to some corrections that would have to be explicitly calculated.
This makes the latter algorithm more attractive and we decided to use it. To improve its performance, we have also used adaptive precision and multiple masses. The former means that the degree of the Chebyshev polynomial is adapted to what accuracy is actually needed in the present iteration step. In practice, this means that when the solver is moving towards the requested precision, the accuracy of approximation can be substantially decreased\footnote{For example, if the degree of Chebyshev polynomial at the start of inversion is typically (for our parameters) of order 150-200, in the final iterations it can go down typically to 30-40 with adaptive precision.}, which saves a factor of around 2 in inversion time.
Multiple mass capability of an algorithm means that for the cost of one inversion for the smallest bare quark mass, one can obtain the solution also for heavier quark masses for practically no additional cost. Since the dependence on the quark mass is central in the present project, the use of a multiple mass variant of the SUMR solver was absolutely essential.

\noindent\textbf{Stochastic sources.} Another important aspect of solving eq. (\ref{dirac_invert}) is the choice of the source $\eta$. The most obvious choice is the point source, which means that the vector $\eta$ is chosen to be 1 at one space-time point and spin-color component and 0 otherwise. This leads to 12 inversions for each gauge configuration, one per spin-color component. However, one can do much better when it comes to statistical error on the pion mass and especially the pion decay constant, with the use of stochastic sources. To keep the signal-to-noise ratio high enough, one has to use timeslice sources, i.e. sources that are non-zero for all spatial points on a given time-slice and for a given spin. This requires 4 inversions per gauge configuration, one per spin component. Moreover, the noise can be further reduced by using the one-end trick, introduced in \cite{oneendtrick}. A reasonable choice of stochastic sources for mesonic correlators are the $\mathcal{Z}(2)$ sources -- the random numbers are of the form $(\pm1\pm i)/\sqrt{2}$. Empirical observations show that this method reduces the statistical error on the pion decay constant by a factor of approximately 2 for the smallest quark masses that we consider, which means that the number of inversions to achieve the same statistical error can be even four times smaller than with point sources.

\section{Setup}
To perform the scaling test of overlap fermions we fix the volume to $L\approx1.3$ fm and use the following ensembles of dynamical $N_f=2$ maximally twisted mass\footnote{Twisted mass (TM) Dirac operator is defined by: $D_{TM}=D_{W}(m)\mathbbm{1}_f+i\mu\gamma_5\tau^3$, where: $m$ -- untwisted quark mass, $\mu$ -- twisted quark mass, $\mathbbm{1}_f$ and $\tau^3$ act in flavour space. For a review of twisted mass fermions, see \cite{shindler}.} \cite{TM}, \cite{TM2} configurations, generated by the European Twisted Mass Collaboration (ETMC) \cite{boucaud}, \cite{boucaud2}:
\begin{itemize}
\item $\beta=3.9$, $V=16^3\times32$, $a\approx0.084$ fm, $a\mu=0.004$, $\kappa=0.160856$,
\item $\beta=4.05$, $V=20^3\times40$, $a\approx0.066$ fm, $a\mu=0.003$, $\kappa=0.157010$,
\item $\beta=4.2$, $V=24^3\times48$, $a\approx0.053$ fm, $a\mu=0.002$, $\kappa=0.154073$.
\end{itemize}
The valence quarks are overlap fermions and the overlap quark mass was chosen to vary from the unitary light quark mass up to the physical strange quark mass. In total we have 20 quark masses, which allows for a precise determination of the quark mass dependence of the pion mass and decay constant.

In addition, we also perform a tree-level scaling test, for which we fix $Nm=0.5$ (which is the equivalent of fixing volume), where $N$ is the number of lattice points in spatial directions and go from $N=4$ to $N=64$. Thus, the change in $N$ introduces the scaling towards the continuum limit, which corresponds to $N\rightarrow\infty$. The temporal extent was set to be 64 times larger than the spatial extent for each value of $N$, which makes it possible to extract the relevant quantities without any contaminations from the excited states. For the details of this test and expressions for the tree-level quark propagators and correlation functions, see \cite{cichy}, \cite{cichy2}.

\section{Results}
\subsection{Tree-level test}
\begin{figure}
\begin{center}
\includegraphics[width=0.5\textwidth,angle=270]{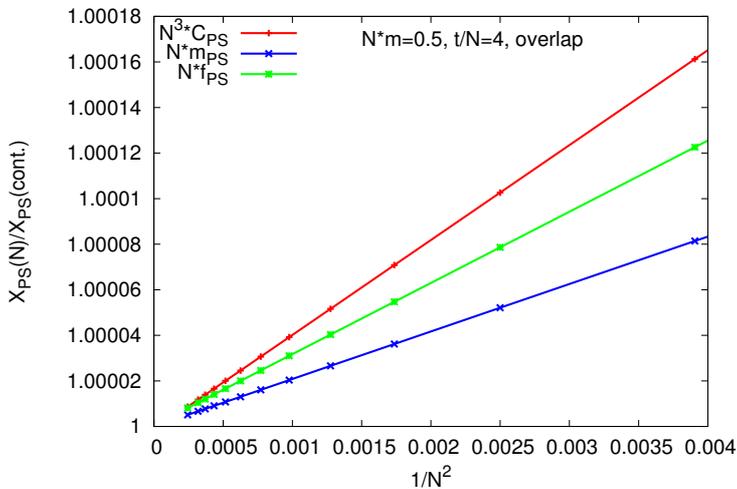}
\caption{The relative cutoff effects of the pseudoscalar correlator at a fixed physical distance, pseudoscalar mass and decay constant. \label{overlap-treelevel}}
\end{center}
\end{figure}

We investigated the relative (with reference to the continuum-limit value) cut-off effects of three observables\footnote{All of the observables are multiplied by an appropriate power of $N$ to make them dimensionless.}: pseudoscalar (PS) correlation function $N^3C_{PS}$ at a fixed physical distance $t/N=4$, pseudoscalar meson mass $Nm_{PS}$ and pseudoscalar decay constant $Nf_{PS}$. The results are presented in Fig. \ref{overlap-treelevel}. As expected, all observables show only $\mathcal{O}(a^2)$ scaling violations, since overlap fermions are $\mathcal{O}(a)$-improved. We also observe that the smallest lattice artifacts are observed for the pseudoscalar mass and the largest for the correlation function itself.

\subsection{The interacting case -- matching the pion mass}
In the interacting case, we are interested in the continuum limit scaling of the pion decay constant for three reference values of the pion mass\footnote{The pion mass and decay constant were obtained from the pseudoscalar correlation function. Thus, pion means here the light pseudoscalar meson.} -- the sea quark pion mass, an intermediate mass $r_0m_\pi\approx1$ and a heavy mass (around the strange quark region) $r_0m_\pi\approx1.5$.\footnote{These two higher reference masses correspond to the partially quenched setup.}

The matching quark mass $\hat m$ is defined, for each 
ensemble, by the condition:
\begin{equation}
  m_\pi^{ov}(\hat m)=m_\pi^{TM}(\mu),
\end{equation}
where $m_\pi^{ov}$ and $m_\pi^{TM}$ are the mixed action and unitary pion 
masses, respectively.

\begin{figure}
\begin{center}
\includegraphics[width=0.5\textwidth,angle=270]{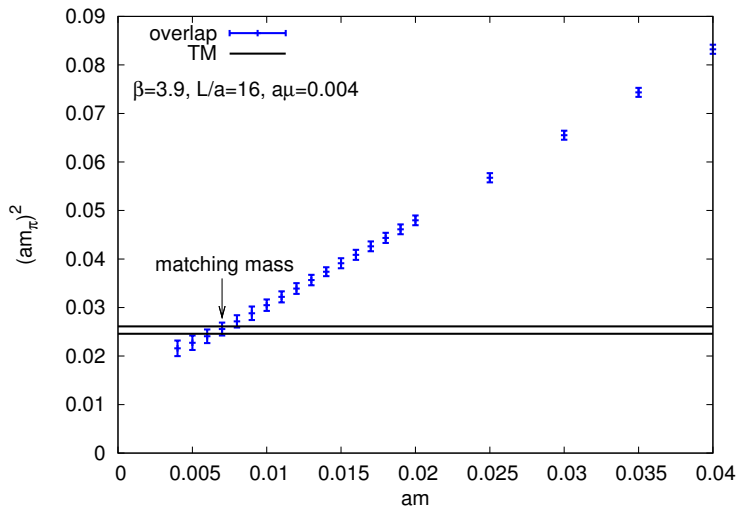}
\caption{Matching mass for ensemble $16^3\times32$, $a\approx0.084$ fm ($\beta=3.9$), $a\mu=0.004$, $\kappa=0.160856$. The matching procedure gives $a\hat m=0.007(1)$. \label{matchingmass}}
\end{center}
\end{figure}

Fig. \ref{matchingmass} shows the procedure of matching for the ensemble at $\beta=3.9$. The ``overlap'' curve shows the dependence of the pion mass on bare overlap quark mass. As the plot indicates, when the overlap quark mass equals 0.007 (with an uncertainty of approx. 0.001), the TM pion and overlap pion have equal masses, corresponding in infinite volume to ca. 300 MeV.
An analogous procedure was applied for other ensembles.

One should add here a word of caution. For the situation of rather small lattice extents of $L\approx1.3$ fm, the chiral zero-modes of the overlap operator can play an important role. Note that since we use here a mixed action approach, these zero-modes are not compensated for by the fermion determinant. The influence of the interplay between the finite box size, the quark mass and the mixed action on the effects of the zero-modes for physical observables will be discussed elsewhere.

\begin{figure}
\begin{center}
\includegraphics[width=0.5\textwidth,angle=270]{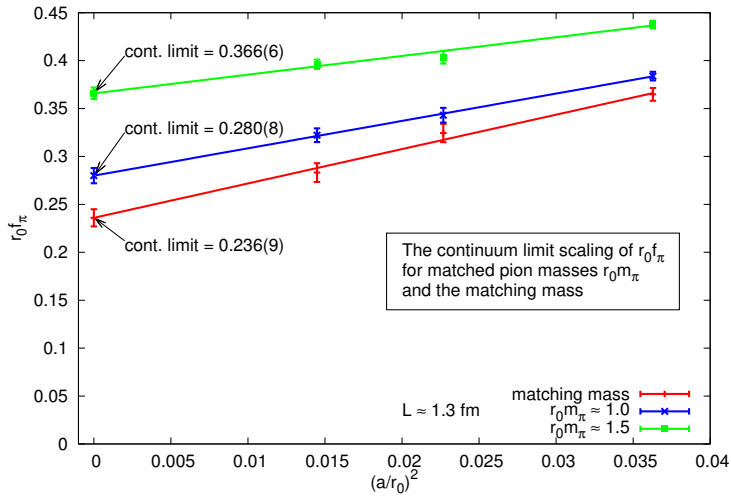}
\caption{Continuum limit scaling of the (overlap) pion decay constant for fixed volume $L\approx1.3$ fm. \label{scaling}}
\end{center}
\end{figure}

\subsection{Continuum limit scaling of the pion decay constant}

Fig. \ref{scaling} shows the continuum limit scaling of $r_0 f_\pi$ for three choices of the pion mass. The lower curve corresponds to the sea quark pion mass, i.e. the bare overlap quark mass is set to the relevant $\hat m$ for each ensemble. For the intermediate and upper curve we fix the pion mass $r_0 m_\pi$ to 1.0 and 1.5, respectively, i.e. we take such overlap quark mass that leads to the chosen value of the pion mass for each ensemble.

We expect $\mathcal{O}(a^2)$ scaling violations and hence we plot the decay constant against the lattice spacing squared.

Indeed, for all three cases, we observe good scaling behaviour towards the continuum limit. The fact that the unitarity violations, proper to a mixed action approach, do not spoil the scaling of the pion decay constant is reassuring.

\section{Conclusion and prospects}
In this paper, we briefly reviewed the overlap discretization of the fermio\-nic action and the techniques we have used to deal with overlap fermions. We presented the continuum limit scaling test of the pion decay constant at tree-level and in the interacting case, using three values of the lattice spacing coming from dynamical twisted mass configurations at roughly matched physical box length of $L\approx1.3$ fm and with three reference values of the pion mass.

We observe very good continuum limit scaling properties for all cases, i.e. at tree-level and in the interacting case for pion masses corresponding to light (with overlap quark mass leading to the same pion mass as the sea quark), intermediate and somewhat heavier quarks.

In the next stage of the project, we plan to extend the scaling test to other observables (e.g. the nucleon mass) and we aim at a detailed study of the particular mixed action setup that we use, including fits of Partially Quenched Chiral Perturbation Theory formulae. We will also investigate the possibility of computing observables for which chiral symmetry is crucial.


\thebibliography{99}
\bibitem{jlqcd} Aoki S. et al., Phys. Rev. D78 (2008), 014508; arXiv:0803.3197.
\bibitem{OVonTM} B\"ar O., Jansen K., Schaefer S., Scorzato L., Shindler A., PoSLAT2006:199,2006; arXiv: hep-lat/0609039.
\bibitem{OVonTM2} Garron N., Scorzato L., PoSLAT2007:083,2007; arXiv: 0710.1582 (hep-lat).
\bibitem{wilson} Wilson K.G., New Phenomena In Subnuclear Physics. Part A. Proceedings of the First Half of the
1975 International School of Subnuclear Physics, Erice, Sicily, July 11 - August 1, 1975, ed.
A. Zichichi, Plenum Press, New York, 1977, p. 69.
\bibitem{nielsen} Nielsen N.B., Ninomiya M., Phys. Lett. B105 (1981), 211.
\bibitem{ginsparg} Ginsparg P., Wilson K., Phys. Rev. D25 (1982), 2649.
\bibitem{neuberger} Neuberger H., Phys. Lett. B 417 (1998), 141; arXiv: hep-lat/9707022.
\bibitem{niedermayer} Niedermayer F., Nucl. Phys. B (Proc. Suppl.) 73 (1999), 105; arXiv:hep-lat/9810026.
\bibitem{hernandez} Hernandez P., Jansen K., L\"uscher M., Nucl.Phys. B552 (1999) 363-378; arXiv:hep-lat/9808010.
\bibitem{luscher} L\"uscher M., Phys. Lett. B428 (1998), 342; arXiv: hep-lat/9802011.
\bibitem{chiarappa} Chiarappa T. et al.; arXiv: hep-lat/0609023.
\bibitem{frommer02} Van den Eshof J. et al., Comput.Phys.Commun. 146 (2002) 203-224; arXiv: hep-lat/0202025.
\bibitem{frommer} Frommer A., Lippert T., Medeke B., Schilling K. (eds.), Numerical Challenges in Lattice Quantum Chromodynamics, Lecture Notes in Computational Science and Engineering 15, Heidelberg 2000.
\bibitem{hyp} Hasenfratz A., Knechtli F., Phys.Rev. D64 (2001), 034504; arXiv:hep-lat/0103029.
\bibitem{oneendtrick} Foster M., Michael C., Phys.Rev. D59 (1999), 074503; arXiv: hep-lat/9810021.
\bibitem{shindler} Shindler A., Phys. Rept. 461 (2008), 37; arXiv: 0707.4093 (hep-lat).
\bibitem{TM} Frezzotti R., Grassi P.A., Sint S., Weisz P., JHEP 0108 (2001) 058; arXiv: hep-lat/0101001.
\bibitem{TM2} Frezzotti R., Rossi G.C., JHEP 0408 (2004) 007; arXiv: hep-lat/0306014.
\bibitem{boucaud} Boucaud P. et al., Phys. Lett. B650 (2007), 304; arXiv: hep-lat/0701012.
\bibitem{boucaud2} Boucaud P. et al., Comput. Phys. Commun. 179 (2008) 695-715; arXiv: 0803.0224 (hep-lat).
\bibitem{cichy} Cichy K., Gonzalez Lopez J., Jansen K., Kujawa A., Shindler A., Nucl.Phys. B800 (2008), 94; arXiv: 0802.3637 (hep-lat).
\bibitem{cichy2} Cichy K., Gonzalez Lopez J., Kujawa A., Acta Phys. Pol. B, Vol. 39 (2008), No.12, 3463-3472; arXiv: 0811.0572 (hep-lat).
\end{document}